**Methodological considerations for estimating policy effects in the context of co-occurring policies**

Beth Ann Griffin, PhD
RAND Corporation, Arlington, VA

Megan S. Schuler, PhD
RAND Corporation, Arlington, VA

Joseph Pane, MSP
RAND Corporation, Pittsburgh, PA

Stephen W. Patrick, MD, PhD
Vanderbilt University Medical Center and School of Medicine, Nashville, TN

Rosanna Smart, PhD
RAND Corporation, Santa Monica, CA

Bradley D. Stein, MD, PhD
RAND Corporation, Pittsburgh, PA

Geoffrey Grimm, MS
RAND Corporation, Arlington, VA

Elizabeth A. Stuart, PhD
Johns Hopkins Bloomberg School of Public Health, Baltimore, MD

**Corresponding Author:**
Beth Ann Griffin
1200 South Hayes Street
Arlington, VA 22202-5050
bethg@rand.org
703-413-1100

**Acknowledgements:**
The authors would like to thank fellow OPTIC team members Rosalie Liccardo Pacula and David Powell as well as members of the OPTIC advisory board, Erin Bagalman, Collen Barry, Richard Frank, Adam Gordon, Karmen Hanson, Keith Humphreys, Christopher Jones, Jeffrey Locke, and Harold Pollack, for their feedback on the initial design and findings from this work. Finally, the authors want to thank Hilary Peterson for her assistance with manuscript preparation and submission.

## Abstract

Understanding how best to estimate state-level policy effects is important, and several unanswered questions remain, particularly about the ability of statistical models to disentangle the effects of concurrently enacted policies. In practice, many policy evaluation studies do not attempt to control for effects of co-occurring policies, and this issue has not received extensive attention in the methodological literature to date. In this study, we utilized Monte Carlo simulations to assess the impact of co-occurring policies on the performance of commonly-used statistical models in state policy evaluations. Simulation conditions varied effect sizes of the co-occurring policies and length of time between policy enactment dates, among other factors. Outcome data (annual state-specific opioid mortality rate per 100,000) were obtained from 1999-2016 National Vital Statistics System (NVSS) Multiple Cause of Death mortality files, thus yielding longitudinal annual state-level data over 18 years from 50 states. When co-occurring policies are ignored (i.e., omitted from the analytic model), our results demonstrated that high relative bias (>85%) arises, particularly when policies are enacted in rapid succession. Moreover, as expected, controlling for all co-occurring policies will effectively mitigate the threat of confounding bias; however, effect estimates may be relatively imprecise (i.e., larger variance) when policies are enacted in near succession. Our findings highlight several key methodological issues regarding co-occurring policies in the context of opioid-policy research yet also generalize more broadly to evaluation of other state-level policies, such as policies related to firearms or COVID-19, showcasing the need to think critically about co-occurring policies that are likely to influence the outcome when specifying analytic models.

**Key Words** concurrent policies; clustered policies; difference-in-differences; state-level policy; policy evaluations; opioid; simulation

# 1. Introduction

Empirical studies that evaluate the effectiveness of a given policy on health outcomes are a staple of health policy research (Angrist & Pischke, 2009). Generally, these studies are observational and capitalize on geographic and temporal variation in policy adoption to identify policy effects (Wing et al., 2018). While it would be ideal from an evaluation standpoint for policies to be adopted in isolation, in practice, jurisdictions often adopt a cluster of policies within a brief span of time which serves to complicate identification of policy effectiveness (Matthay et al., 2021b). Concurrently enacted policies lack sufficient periods of time between enactment dates, posing challenges to isolating the effect of the primary policy independent of co-occurring secondary policies. Recent work has begun to elucidate the methodological challenges faced when trying to disentangle the individual policy effects of concurrent policies (Matthay et al., 2021a); however, unanswered questions remain regarding the magnitude of the impact of co-occurring policies on the performance of commonly used models for estimating policy effects.

Methodologic challenges arising from concurrent policies apply broadly to policy research and are particularly relevant for analyses of opioid policies as states have generally enacted multiple opioid-related policies as the opioid crisis has evolved. For example, many states implemented some combination of naloxone laws, Good Samaritan laws, and medical marijuana laws during 2015-2017. In addition to pre-existing prescription drug monitoring program (PDMP) laws, by 2017 the majority of states had implemented at least 3 of these 4 categories of policies. Thus, analytically, it is very unlikely that it will be possible to identify a sizeable number of states that have implemented the policy of interest in "isolation." Furthermore, these policies and regulations may be enacted in rapid succession. For example, Florida nearly simultaneously enacted a PDMP and stricter pain clinic regulations (Kennedy-Hendricks et al., 2016; Rutkow et al., 2015). Similarly, as opioid-related overdoses climbed, many states adopted naloxone access laws and Good Samaritan laws simultaneously or in short succession to reduce opioid overdose mortality (Abouk et al., 2019; Blanchard et al., 2018; McClellan et al., 2018).

In a prior scoping review of evaluation studies examining the impact of U.S. state- and federal-level policies on opioid-related outcomes published in 2005-2018, it was found that of the 145 studies reviewed, 94 (65%) did not control for any co-occurring policies (Schuler et al., 2020). Studies published more recently were more likely to adjust for co-occurring policies, likely reflecting both the increasing number of state policies enacted, as well as increased attention to this methodological concern, which was highlighted in the review. When co-occurring policies were accounted for, the set of polic(ies) adjusted for varied notably across studies. For example, comparison of two published PDMP studies found that both control for a fully disjointed set of policies: pill mill laws, ACA Medicaid expansion, OxyContin reformulation, and passage of Medicare Part D (Mallatt, 2018) compared to naloxone laws, Good Samaritan laws, requirement for physical examination for prescribing and identification upon dispensing, and medical marijuana laws (Dave et al., 2018). However, few studies explicitly discussed the rationale underlying the co-occurring policies they adjusted for. More broadly (Matthay et al., 2021a) found that only a third of social policy studies explicitly considered the presence of concurrent policies.

In practice, policy evaluation studies are more likely to account for potential confounding due to differences in state-level characteristics (e.g., sociodemographic characteristics) than confounding due to differences in state policy environments (Schuler et al., 2020). However, co-occurring policies could represent a more significant confounding threat, as these policies may have a stronger relationship with the outcomes compared to other state-level confounding factors. As such, understanding the potential magnitude of the bias that might be due to confounding from co-occurring policies in state-level policy evaluation studies could enhance the rigor of policy analyses and policymaking. Obtaining unbiased estimates of policy effects is imperative to identifying and implementing effective policies that can improve public health and well-being across a range of areas, including opioid misuse as well as gun violence and COVID-19 mitigation.

In this paper, we conducted a simulation study to examine the impact of co-occurring policies on the performance of commonly used models in opioid policy evaluations. We had two key objectives – (1) in the context of correctly specified models that control for co-occurring policies, we sought to determine the minimum length of time between enactment of the primary and co-occurring policies needed to obtain unbiased effect estimates, and (2) to determine the impact of model misspecification that omits the co-occurring policy on estimation of the effect of the primary policy. Simulation conditions vary effect sizes of the primary and co-occurring policies as well as the length of time between enactment dates of the primary and co-occurring policies, among other factors. The simulations focused on evaluations utilizing longitudinal state-level ecological data or panel data, not on the use of individual or county level data nested within states. The results provide insights and best practice guidelines for applied policy researchers.

## 2. Methods

We utilized Monte Carlo simulations to assess the effect of concurrent policy enactment on the evaluation of state-level policies. We considered the performance of two different statistical models – a *correctly specified model* that includes regression terms for both policies simultaneously and a *misspecified model* that omits the co-occurring policy – over a range of scenarios where a group of "treated" states enacted two policies (i.e., the primary policy and a co-occurring, secondary policy). In the context of the correctly specified model, we examined simulation conditions varying the length of time and relative timing of the policy enactment dates for the two policies in order to identify the minimum length of time needed between policy enactment dates to obtain robust estimates of the primary policy. Additionally, we explored the impact of misspecifying the statistical model by omitting the co-occurring policy on estimation of the primary policy effect. The corresponding author's Institutional Review Board approved this study.

### 2.1 Data

The simulation was based on longitudinal annual state-level data. The outcome of interest was the annual state-specific opioid mortality rate per 100,000 state residents, obtained from the 1999-2016 National Vital Statistics System (NVSS) Multiple Cause of Death mortality files. Eighteen years of annual observations across 50 states provided a total of 900 observations. Consistent with other studies (Abouk et al., 2019; Chan et al., 2020; Kilby, 2015), opioid-related overdose deaths were identified based on ICD-10-CM-external cause of injury codes indicating accidental and intentional poisoning (X40-X44, X60-64, X85, Y10-Y14) in combination with an opioid diagnosis codes (T40.1 poisoning by heroin, T40.2 poisoning by natural and semisynthetic opioids, T40.3 poisoning by methadone, and T40.4 poisoning by synthetic opioids excluding methadone). To avoid concerns about model overfitting, we included a single covariate: annual state-level unemployment rate, derived from the U.S. Department of Labor, Bureau of Labor Statistics (U.S. Department of Labor, 2019). This covariate was selected because of the frequency of its use in opioid policy studies (Schuler et al., 2020). In our simulation, the original opioid mortality data represented outcomes under the null treatment condition (i.e., no policies in effect for a given state and given year); outcomes reflecting the effect of simulated primary and secondary policies were generated for treated states as described below.

*2.2 Data Generation*

The simulation design builds directly from similar prior work that compared statistical methods for evaluating the impact of state laws on firearms deaths (Schell et al., 2019) and opioid-related mortality (Griffin et al., 2021). In each simulation iteration, we first selected a random subset of $k$ states to be the "treated" group that enacted both the primary and co-occurring policy, with the remaining states serving as the comparison group that never enacted either policy. Two time-varying policy indicators were generated denoting whether the primary policy ($A_{1it}$) and co-occurring policy ($A_{2it}$) were in effect for each state and each year. For comparison states, $A_{1it} = A_{2it} = 0$ for the entire study period.

Our study period included 18 years of repeated time series data, spanning 1999 to 2016. Across simulation conditions, we varied the length of time between the primary and co-occurring policy enactment dates considering 4 conditions: approximately 0-1 years apart, 3-4 years apart, 6-7 years apart, and 9-10 years apart. For each policy state, policy enactment dates were selected as random draws from a multivariate normal distribution where the mean for the first policy was defined as a randomly selected year between 1999 and an upper bound that ensured there was enough time for the second policy to occur (e.g., 2007 for the 9-10 year interval) and where the mean for the second policy was defined as the previously drawn enactment date for the first policy plus the minimum length of the specified interval (e.g., 9 years for the 9-10 year interval). The variances for the multivariate normal distribution were set equal to 1 and the covariance was set equal to 0.9. This process yielded two continuous enactment dates (e.g., 1999.591 and 2000.177 for a sample draw when the policies where to be enacted right on top of each other), which we then rounded to the nearest month and year of enactment for the two policies. Of note, in the first year of enactment, $A_{1it}$ and $A_{2it}$ were coded as fractional values between 0 and 1, indicating the percentage of the year each policy was in effect. Once the primary or secondary policy was enacted, they remained in effect throughout the study period; thus, $A_{kit} = 1$ (for $k$=1,2) for all remaining years.

We generated the outcome values ($Y_{it}^*$) using the following formulas:

$$Y_{it}^* = Y_{it} + te_1 * A_{1it} + te_2 * A_{2it}$$

where $Y_{it}$ denotes the observed outcome value for state $i$ in year $t$ (obtained as described in the Data section) and $te_1$ and $te_2$ denote the policy effects for the primary and co-occuring policies. We note that our simulation study focused on conditions in which the policy effect was homogenous across states and did not vary over time.

Simulation conditions also varied the following factors:

*(1) Policy effect size.* We considered eight different treatment effect scenarios, namely ($te_1, te_2$) = (0%, 0%) (e.g., both policies had null effects), (0%, -15%), (-15%, 0%) (-10%, -10%), (-15%, -5%), (-5%, -15%), (-10%, -20%), and (-20%, -10%). These treatment effects – meant to cover a range of settings of realistic effect sizes of magnitudes that would be of interest to policymakers -- were selected after consultation with our research center's advisory board, which is comprised of senior researchers active in opioid or substance use disorder policy research, former and current policymakers, and representatives from policymaking organizations.

*(2) Number of treated states.* We also investigated the role of sample size, considering two conditions in which both policies were enacted by 5 and 30 states, respectively.

(3) *Ordering of policy enactment dates.* Finally, we considered cases where the order of enactment was random (e.g., ~50% of the time the primary policy would occur first) and cases where the order of enactment was fixed so the primary policy always preceded the secondary policy.

*(4) Form of policy effect. W*e considered two possible ways in which the policies have an effect on the outcome: an instantaneous effect and a 3-year linear phase-in effect. We specified an instantaneous effect as a simple step-function that has a value of one when the policy in question is in effect, zero otherwise. The 3-year linear phase-in policy effect allowed for the effect of a given policy to grow linearly in the first 3 years after implementation, with values starting at zero and reaching 1 after 3 years of implementation.

(5) *Varying the relative timing of the first policy:* Our simulation design easily allows us to control for when in the time series the enactment of the first policy occurs. Thus, we explored (for a subset of our simulations) the effect on performance of the models when the timing of the first policy was restricted to occur at the beginning (1999-2004), middle (2005-2011) or end of our time series (2011-2016).

(6) *Having a subset of states that only enact a single policy.* Since, in practice, not all states with the primary policy will have also enacted the co-occurring secondary policy, we explored the effect of having a subset that only enacts a single policy in two ways. First, we examined the effect of adding a subset of states that enacted a single policy to settings with 30 states enacting both policies. Here we explored the addition of 5 and 10 states per policy, hypothesizing that performance will improve due to the increased sample size. Second, we examined the effect of swapping a subset of our 5 and 30 states to have only a single policy instead of both policies.

For each permutation, we assessed performance across 5000 randomly generated datasets. We conducted all simulations in R, using the OPTIC.simRM package which implements our simulation code on user-provided outcome data. The package is currently available on github https://github.com/aescherling/optic-core, including code to create the figures. Extensive results for all statistical models considered in our simulation are available via a Shiny tool (www.rand.org/t/TLA1975-1).

*2.3 Candidate Statistical Models*

We focused on the performance of a linear autoregressive (AR) model as this model was recently identified as an optimal model for estimating the effects of a single state-level opioid policy on opioid-related mortality (Griffin et al., 2021). Given the continued popularity of the classic two-way fixed effects (FE) model (commonly used in context of difference-in-differences [DID] design), we additionally provide selected results for the two-way FE model.

We note again that our design seeks to determine how the AR and classic two-way FE model handle concurrent (e.g., confounding) policies under the simple scenario where we have homogenous treatment effects. As has been noted in the recent literature, the classic two-way FE model will be biased in the presence of heterogeneous treatment effects (Callaway & Sant'Anna, 2021; Goodman-Bacon, 2021); however, this concern is not applicable in our simulation context as we have homogenous treatment effects.

For both the AR and two-way FE models, we examined two model specifications. First, we fit the *correctly specified version* of the outcome model that includes regression terms for both the primary and the co-occurring policy; second, we fit a *misspecified version* of the outcome model that omits the co-occurring policy.

The correctly specified AR model was:

$$Y_{it}^* = \alpha_1 \cdot (A_{1it} - A_{1i,t-1}) + \alpha_2 \cdot (A_{2it} - A_{2i,t-1}) + \beta \cdot X_{it} + \gamma \cdot Y_{it-1}^* + \sigma_t + \epsilon_{it} \qquad (1)$$

where $X_{it}$ denotes the time-varying state-level covariate (unemployment rate) and $\varepsilon_{it}$ denotes the error term. Notably, Model (1) includes time FE, $\sigma_t$, to quantify temporal trends across time, but adjusts for state-specific variability with the AR term ($\gamma \cdot Y_{it-1}^{obs}$) rather than state FE. Inclusion of the AR term created a "change" model, as the policy effect was defined as the expected difference in the outcome, given the prior year's outcome. As such, we coded both policy variables using *change coding* – i.e., $(A_{1it} - A_{1i,t-1})$ – based on work demonstrating that effect size estimates from AR models can be substantially biased when using standard *effect coding* (i.e., $A_{it}$) (Cochrane & Orcutt, 1949). The coefficient estimates ($\hat{\alpha}_1, \hat{\alpha}_2$) on the change coding terms represented the AR estimator of the policy effects of $(A_1, A_2)$. For this work $\hat{\alpha}_1$ was of primary interest. However, results additionally included model estimates for the effect of the co-occurring policy $(\hat{\alpha}_2)$.

The misspecified AR model was:

$$Y_{it}^* = \alpha_{1M} \cdot (A_{1it} - A_{1i,t-1}) + \beta \cdot X_{it} + \gamma \cdot Y_{it-1}^* + \sigma_t + \epsilon_{it} \qquad (2)$$

where the co-occurring policy variable was omitted from the model. Interest was in understanding the resulting bias in the estimate of the primary policy under the misspecified model and how this varied with the relative timing of the primary and co-occurring policies. Model (2) is often fit in practice when researchers focus on a single policy of interest without controlling for any additional co-occurring policies as potential confounders (e.g., (Kuo et al., 2016; Pauly et al., 2018; Phillips & Gazmararian, 2017; Yarbrough, 2018). For the misspecified model, only estimates of the primary policy effect $(\hat{\alpha}_{1M})$ are reported.

In contrast, the classic two-way FE model has the following version of the correctly specified and misspecified models, respectively:

$$Y_{it}^* = \alpha_1 \cdot A_{1it} + \alpha_2 \cdot A_{2it} + \beta \cdot X_{it} + \eta_i + \sigma_t + \epsilon_{it} \qquad (3)$$

$$Y_{it}^* = \alpha_1 \cdot A_{1it} + \beta \cdot X_{it} + \eta_i + \sigma_t + \epsilon_{it} \qquad (4)$$

where equations (3) and (4) now include state-level fixed effects, $\eta_i$, for state $i$ and where the policy exposure variables are captured using classic effect coding $(A_{kit})$ for $k$ = 1 and 2.

All models used state population as an analytic weight, an approach commonly used in opioid policy evaluations (e.g., (Ali et al., 2017; Buchmueller & Carey, 2018; McInerney, 2017; Paulozzi et al., 2011).

*2.4 Metrics for Assessing Relative Performance of Candidate Statistical Methods*

We report two primary statistical metrics commonly used to judge model performance with respect to both the estimated effect of the primary policy ($\widehat{\alpha}_1$ and $\widehat{\alpha}_{1M}$) (from both Models (1) and (2)) and the estimated effect of the co-occurring policy ($\widehat{\alpha}_2$) (from Model (1)).

*(1) Bias.* Bias assesses the average difference between the estimated effect and true effect across all simulations. It indicates the tendency of the estimated effects of a given model to fall closer or further from the true effect on average. For the null effect settings, bias is shown on an effect size scale to denote small (<0.2), moderate (0.2-0.4) and large (>0.4) bias. For the non-null settings, the percent relative bias is shown, with <5% denoting virtually no bias, 5-10% small bias, 10-20% moderate bias and >20% large bias (Li & Li, 2021). Since all of our assumed effects were negative, positive relative bias implied that the estimated effect was overestimated in the negative direction, while negative relative bias implied that the estimated effect was underestimated and closer to the null, relative to the true effect.

*(2) Variance.* Variance is calculated as the average variance of the estimated effect across all simulations and indicates the precision of the estimated effects of a given model.

Additional metrics can be found in our Shiny app. These included Type I and S errors and coverage rates, given the high prevalence of frequentist null hypothesis significance testing in the state-level policy evaluation literature.

**3. Results**

*3.1 Correctly specified AR and two-way FE models: Increasing length of time between primary and co-occurring policy enactment*

For the AR model, we found that the length of time between enactment dates had minimal effect on relative bias for the correctly specific model: across all 4 interval lengths considered, the relative bias was always minimal (< 2%) (**Figure 1**). However, the variance was notably larger for the 0-1 year interval compared to longer intervals. The variance remained relatively stable once the length of time between the enactment dates was longer than 3-4 years. These same trends were observed for the two-way FE model. Comparing the AR and two-way FE models, the AR model generally yielded smaller relative bias and lower variance than the two-way FE model, suggesting the AR model specification is more robust to decreasing the length of time between enactment dates than is the two-way FE model.

*3.2 Misspecified AR and two-way FE models: Increasing length of time between primary and co-occurring policy enactment*

We next consider performance of misspecified models that omit the co-occurring policy term in the model. As expected, the relative bias for the estimated primary policy effect was larger in magnitude for the misspecified models (**Figure 2**) compared to the correctly specified models (**Figure 1**). For the AR model, the relative bias was quite large (82%) when there was only 0-1 years between enactment dates. Notably, relative bias was much smaller (approximately -7%) and similar across conditions in which the length of time between enactment dates was 3-4-years or greater. Variance was relatively unaffected by the length of time between enactment dates in these AR models.

Comparing the performance of the misspecified AR and two-way FE models, we see that the two-way FE model yielded notably biased effect estimates when the enactment dates were 0-1 years apart (90%) or 3-4 years apart (45%). Additionally, the variance of the policy effects for the two-way FE models was approximately three times higher than the variance for the AR models.

*3.3 Correctly specified AR model: Effect of treatment effect size and policy ordering*

Notably, relative bias was consistently higher in the simulation setting in which the primary policy always was enacted first ("ordered" panel in **Figure 3**) compared to the setting in which the ordering of the two policies was randomly determined ("unordered" panel in **Figure 3**). The relative bias in the ordered setting ranged from 11% to 23% for the primary policy and from -6% to -24% for the co-occurring policy, whereas estimates from the unordered setting consistently had minimal relative bias. Additionally, in the ordered setting, the models consistently overestimated the effect of the primary policy and underestimated the effect of the co-occurring policy, which was enacted second, showcasing how the model attributed some of the effect of the (second) co-occurring policy to the primary policy.

Critically, in the ordered setting, relative bias of both primary and co-occurring policies was related to effect size, with greater bias observed when the true policy effect was smaller. In cases where the true effect of the primary or secondary policy was 5%, we saw evidence that relative bias could be as high as 25-30%. In contrast, in cases where the true effect of the primary or secondary policy was 20%, relative bias would shrink to around 5%. In the unordered setting, bias was minimal regardless of effect size (always <5%).

*3.4 Correctly specified AR model: Comparing phased-in policy effects to immediate policy effects*

Performance was consistently worse for our models when the form of the policy effects was assumed to have a 3-year phase-in period (**Figure 4**) compared to conditions in which the policy had an immediate effect (as shown in **Figure 1**). We note that the policy variables used in the models always accurately reflected the form of the policy used in the data generation (e.g., for the linear phase-in condition, the policy variable was coded as a proportion (reaching 1 at the end of the 3 year phase-in). Decreased performance (relative to the immediate effect conditions) is observed because there are fewer years for which the primary policy is at "full effect." Specifically, relative to conditions in which the policy had an immediate effect, bias for both the primary and secondary policies was larger (ranging from ~0% to 12% across all scenarios) and variance was larger (8-fold when the policies were enacted within 0-1 years of each other; 2-fold for all other lengths in difference in enactment dates). While bias remained constant across scenarios, variance notably improved when at least 3 years were between the enactment dates of the co-occurring policies. As with instantaneous models, performance is also weakened when the policies are ordered (see Shiny app for full results).

*3.5 Correctly specified and misspecified AR models: Effect of relative timing of enactment of first policy*

The bias and variance were greatest when the timing of the enactment for the first policy occurs earlier versus later in the time series, though bias for all scenarios continued to be low in the correctly specified models (e.g., < 7.5%; **Figure 5**). Variance also decreased for the estimated policy effects in the settings where the policies occurred later in the time series (see Shiny app).

Overall performance was best for the AR models shown in **Figure 1** and **2** where enactment dates were allowed to uniformly occur over the fuller length of the time series. Notably, bias in the misspecified models was higher for the models where the first policy was enacted earlier in the time series (>95%) versus the end of the time series (~82%) with bias minimized when the enactment dates were allowed to occur uniformly over the time series as shown in **Figure 2**.

*3.6 Correctly specified AR model: Inclusion of states that only enacted a single policy*

As shown, adding additional states that only enact a single policy to our scenarios improved performance of a given model by decreasing bias and variance in the estimated effects of the primary and secondary policies (**Figure 6**). In contrast, there was a slight decrease in precision for both policy effects when we swapped states enacting policies in such a way that only a subset enacts a single policy, which presumably was because of the resulting decrease in the total number of policy enacting states used to estimate the policy effects. There was little influence on bias in these scenarios.

## 4. Discussion

Overall, we conducted a novel simulation study to (1) assess the minimum length of time between enactment of the primary and co-occurring policies needed to obtain unbiased effect estimates and (2) to determine the impact that model misspecification – namely, omitting the co-occurring policy – has on estimation of the primary policy effect. We note that all simulations focused on the case of homogenous treatment effects that do not vary over time. As we discuss below, we found that the required interval between primary and co-occurring policies necessary to obtain robust policy estimates varied across model specifications but was generally shorter for AR models compared to two-way FE models. Additionally, our results demonstrated that the magnitude of the bias may be quite substantial (e.g., 60%) in some contexts when confounding co-occurring policies are omitted from the analytic model. As such, researchers should think carefully about co-occurring policies that are likely to influence the outcome and ensure inclusion of them in analytic models. Our findings highlight several key issues regarding co-occurring policies in the context of opioid-policy research, yet also generalize more broadly to evaluation of other state-level policies, such as policies related to firearms or COVID-19.

This study aimed to identify the minimum interval needed between enactment of two policies to obtain unbiased estimates of the primary policy. Our findings indicated that AR models that adjusted for co-occurring policies were able to obtain accurate policy effect estimates even when the policies were enacted in rapid succession (0-1 years) but yielded more precise estimates when policies were at least 3-4 years apart. Notably, AR models outperformed classic two-way FE models on this front, as two-way FE models yielded estimates with much greater variance under all scenarios. Thus, our findings demonstrate that controlling for co-occurring policies will effectively mitigate confounding bias; however, effect estimates may be relatively imprecise (i.e., with larger variance) when policies are enacted in near succession. The relative imprecision of policy estimates in this setting may result in spurious null findings, in which policies are deemed to be ineffective when they truly are effective.

These findings inform important guidance for applied researchers conducting policy evaluation studies in the context of concurrently enacted policies. Specifically, we strongly encourage researchers to carefully examine the distribution of times between enactment dates for policies being examined – e.g., by computing the length of time between enactment dates for each state.

Examining the distribution of length of time between enactment dates across states will be informative regarding the potential for bias in the estimated effects of the primary and co-occurring policies. If there is sufficient time between the enactment dates of the policies (e.g., at least 3-4 years for AR model, at least 6-7 for two-way FE model), a regression model that controls for all co-occurring policies will have minimal to no confounding. In cases when the policies were enacted very closely together, results from our study and simulation tool can be used to gauge the potential size of the bias for a particular analysis.

Researchers should also be sure to make a note about the ordering of the co-occurring policies. Performance was notably worse when the primary policy always preceded the secondary policies, particularly when the policies were enacted in close succession. This happens frequently in practice. For example, in existing research on cannabis policies, recreational legalization of cannabis has always occurred after the enactment of state medical cannabis legalization (Pacula & Smart, 2017), creating the potential for more bias on the primary policy effects when those policies are enacted close together.

Our findings also demonstrate that, in cases where the co-occurring policy is a true confounder, omitting it from the regression model will introduce confounding bias into the estimate of the primary policy. In this scenario, the effect of the co-occurring policy was essentially attributed to the primary policy. Under simulation conditions that mimic realistic settings in opioid policy research, our results indicated that the magnitude the bias for estimate of the primary policy may be as large as 82% when co-occurring policies are enacted within rapid succession of each other (0-1 years of each other). We note that in our simulation study, we generated a single co-occurring policy; in reality, it is likely that there may be multiple co-occurring policies. Conceptually, our co-occurring policy can be thought of as the cumulative effect of all co-occurring policies; as such, our findings could be interpreted as an upper-bound. Specifically, in practice, controlling for some, but not all, co-occurring policies could be expected to attenuate bias relative to our results from the misspecified model, assuming that all of these policies had effects in the same direction. We note that additional co-occurring policies enacted at different times, with varying confounding directions, creates a more complex scenario regarding variation in the length of overlap between the primary and different co-occurring policies, which may result in time-varying confounding or moderating effects between the policies themselves. As such, the complexities of multiple, time-varying co-occurring policies is an important area for ongoing research. In the setting of multiple co-occurring policies, we recommend a diagnostic step of examining the distribution of length of time between policies for all pairwise comparisons. This will allow researchers to identify for which sets of policies it might be difficult to disentangle the effects of. In practice, the strength of confounding likely varies across co-occurring policies; as we discuss below, researchers should prioritize adjustment for policies anticipated to have the strongest confounding effect.

We underscore that due to the interconnected nature of the opioid ecosystem, co-occurring policies do not need to directly target the outcome being examined to still have a confounding effect on the estimated effect of the primary policy. For example, while mandatory PDMPs are an important co-occurring policy that might confound studies of prescribing guidelines, as both directly impact prescribing (Castillo-Carniglia et al., 2021; Encinosa et al., 2021; Mauri et al., 2020; Puac-Polanco et al., 2020), such PDMPs may also be a confounder for naloxone policies examining overdose since mandatory access PDMPs have also been shown to indirectly have a protective effect on overdose rates (Ji et al., 2021; Lee et al., 2021; Puac-Polanco et al., 2020). However, not all co-occurring policies are likely confounders. For example, policies likely to increase the number of buprenorphine waivered prescribers, such as hub-and-spoke policies (Snell-Rood et al., 2020), would likely have little effect on opioid analgesic prescribing.

Additionally, our findings indicate that the correctly specified model was notably robust in all scenarios considered when fitting a linear AR model to opioid-related mortality when there is variation in the ordering of enactment of the primary and co-occurring policies across states. However, performance deteriorates meaningfully in the cases where the primary policy always comes first and/or when the length of time between the two policies is short (0-1 years; relative bias >20% for the primary policy). Intuitively this makes sense, as fixed ordering between policies essentially provides less information because the policy enacted second is never observed in the absence of the first policy. In practice, it is likely that there will be variation in policy ordering across states; however, in some contexts, adoption of one policy may consistently precede adoption of another policy (especially when the number of treated states is small).

The study has several important limitations that should be considered alongside the findings. Our simulation was relatively simplistic, representing a reasonable starting point for studying model performance with concurrently enacted policies. Future work should consider additional nuances that might impact our findings and provide additional guidance to the field. For example, it would be meaningful to explore the ability of the models to identify potential synergistic effects between the co-occurring policies and to consider additional outcomes. Additionally, it would be of interest to examine whether the use of lower-level data (e.g., individual-level or county-level data) can reduce bias and increase precision of state-level policy evaluations involving more than one policy. Finally, we note that we only considered one potential form for time-varying effects of our primary and secondary policies -- namely a linear phase-in period over three years. We note that all of the settings considered in this work assume homogeneous treatment effects across states. The general policy evaluation methods field is rich with calls for more robust handling of heterogeneous time-varying effects of policies (Callaway & Sant'Anna, 2021; Goodman-Bacon, 2021; Sun & Abraham, 2021), yet there is a notable dearth of methods available for capturing the time-varying effects of multiple policies, with a notable exception being the recent working paper by De Chaisemartin and d'Haultfoeuille (2022) and Schell et al. (2020). Future research is needed to both expand current staggered adoption methods for time-varying policy effects to handle more than one policy as well to better understand how all of these methods compare in the face of different types of time-varying effects and time-varying confounding.

## Declarations

**Funding:** This research was financially supported through a National Institutes of Health (NIH) grant (P50DA046351, PI: Stein). NIH had no role in the design of the study, analysis, and interpretation of data nor in writing the manuscript.

**Conflicts of interest/Competing interests:** The authors have no relevant financial or non-financial interests to disclose.

**Availability of data and material:** The data that support the findings of this study are available from National Vital Statistics System (NVSS) Multiple Cause of Death mortality files (1989 through 2018) but restrictions apply to the availability of these data, which were used under license for the current study, and so are not publicly available. The data can be request under a similar license (data use agreement) from the NVSS.

**Authors' contributions:** The authors jointly conceived the idea for the study. BG and RS derived the first iteration of the design of the simulation. BG and GG designed, developed and implemented the original simulation code; JP lead all adaptions and runs under consultation with all authors. GG and JP developed needed graphics and the associated Shiny app for this work. BG, MS, and ES drafted the manuscript with input from all authors. All authors extensively edited and provided input on all phases of the study and all authors read and approved the final manuscript.

**Ethics approval:** All procedures performed in studies involving human participants were in accordance with the ethical standards of the institutional and/or national research committee and with the 1964 Helsinki Declaration and its later amendments or comparable ethical standards. The corresponding author's Institutional Review Board deemed this study exempt (Human Subjects Assurance Number 00003425 (6/22/2023).

**Consent to participate:** Not applicable

**Consent for publication:** Not applicable

**Figure Legends**

Fig. 1 Model performance for correctly-specified AR and two-way FE models

Note:

Simulation condition:

- 30 treated states

- Both policies decrease opioid-related mortality by 10%

- Both policies have an instantaneous effect

- Policies are randomly ordered

Fig. 2 Model performance for misspecified AR and two-way FE models that omit co-occurring policy term

Note:

Simulation condition:

- 30 treated states

- Both policies decrease opioid-related mortality by 10%

- Both policies have an instantaneous effect

- Policies are randomly ordered

Fig. 3 Model performance for the correctly specified linear AR model as a function of policy effect sizes and policy ordering

Note:

Simulation condition:

- 30 treated states

- Policies were enacted within 0-1 years

- Both policies have an instantaneous effect

Fig. 4 Model performance for correctly-specified AR models when policy has a linear phase-in effect of 3 years

Note:

Simulation condition:

- 30 treated states

- Both policies decrease opioid-related mortality by 10%

- Both policies have a 3-year linear phase-in effect

- Policies are randomly ordered

Fig. 5 Model performance for AR models as a function of the relative timing of the first enacted policy in the time series

Note:

Simulation condition:

- 30 treated states

- Both policies decrease opioid-related mortality by 10%

- Both policies have an instantaneous effect

- Policies are randomly ordered

Fig. 6 Model performance for correctly specified AR models when a subset of states only implemented a single policy

Note:

Simulation condition:

- Both policies decrease opioid-related mortality by 10%

- Both policies have an instantaneous effect

- Policies are randomly ordered

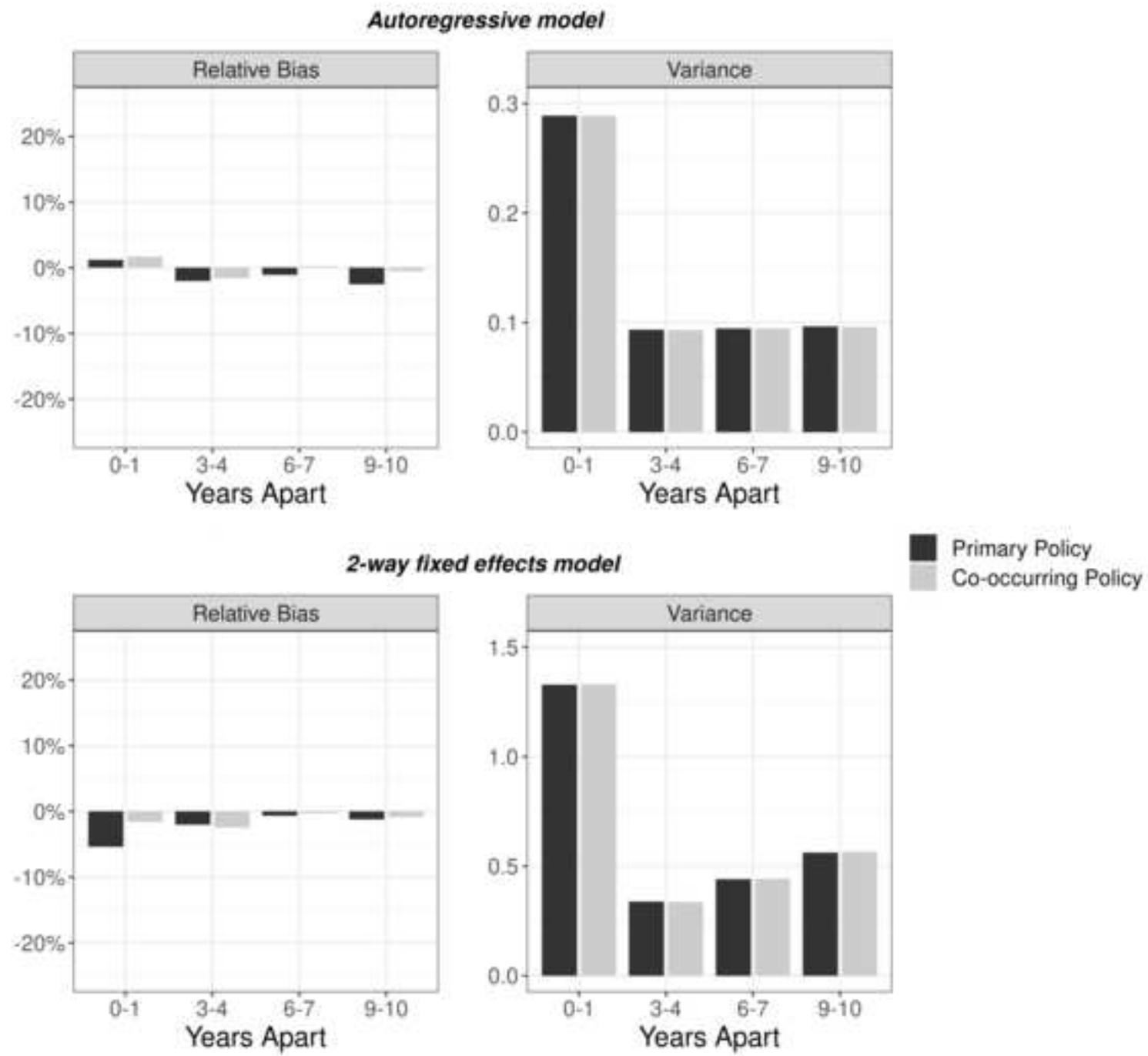

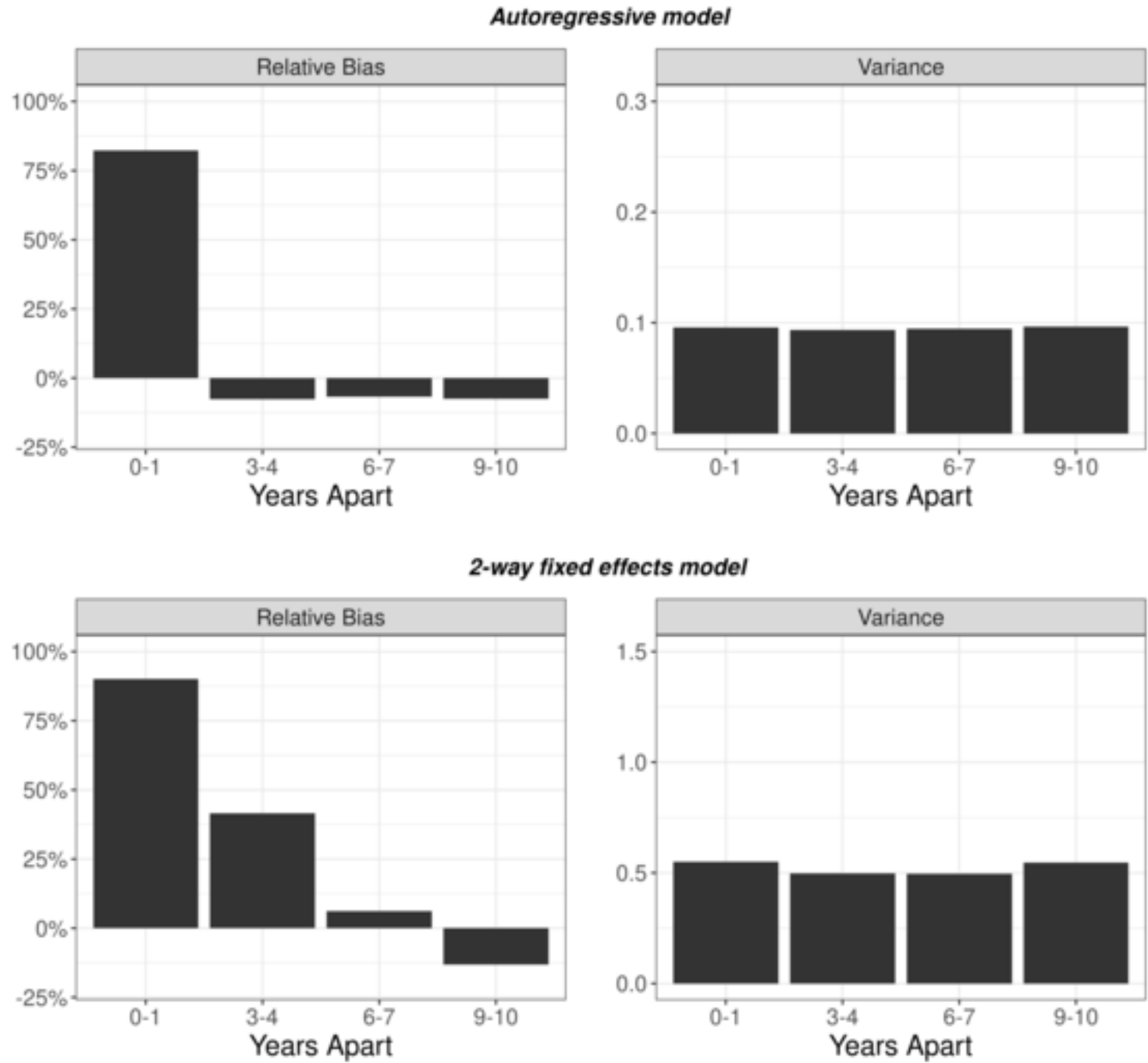

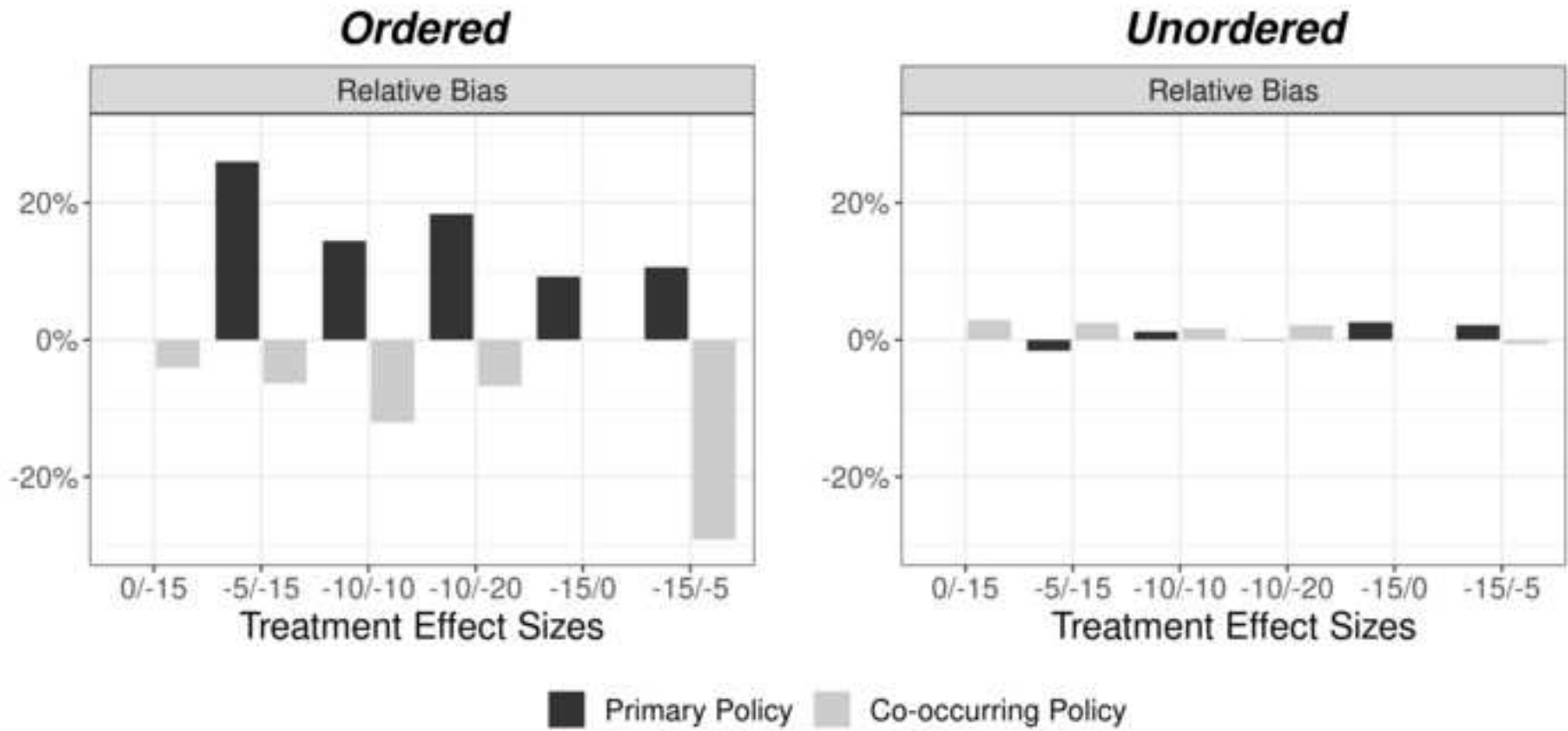

Note: Relative bias is undefined when the effect size is 0.

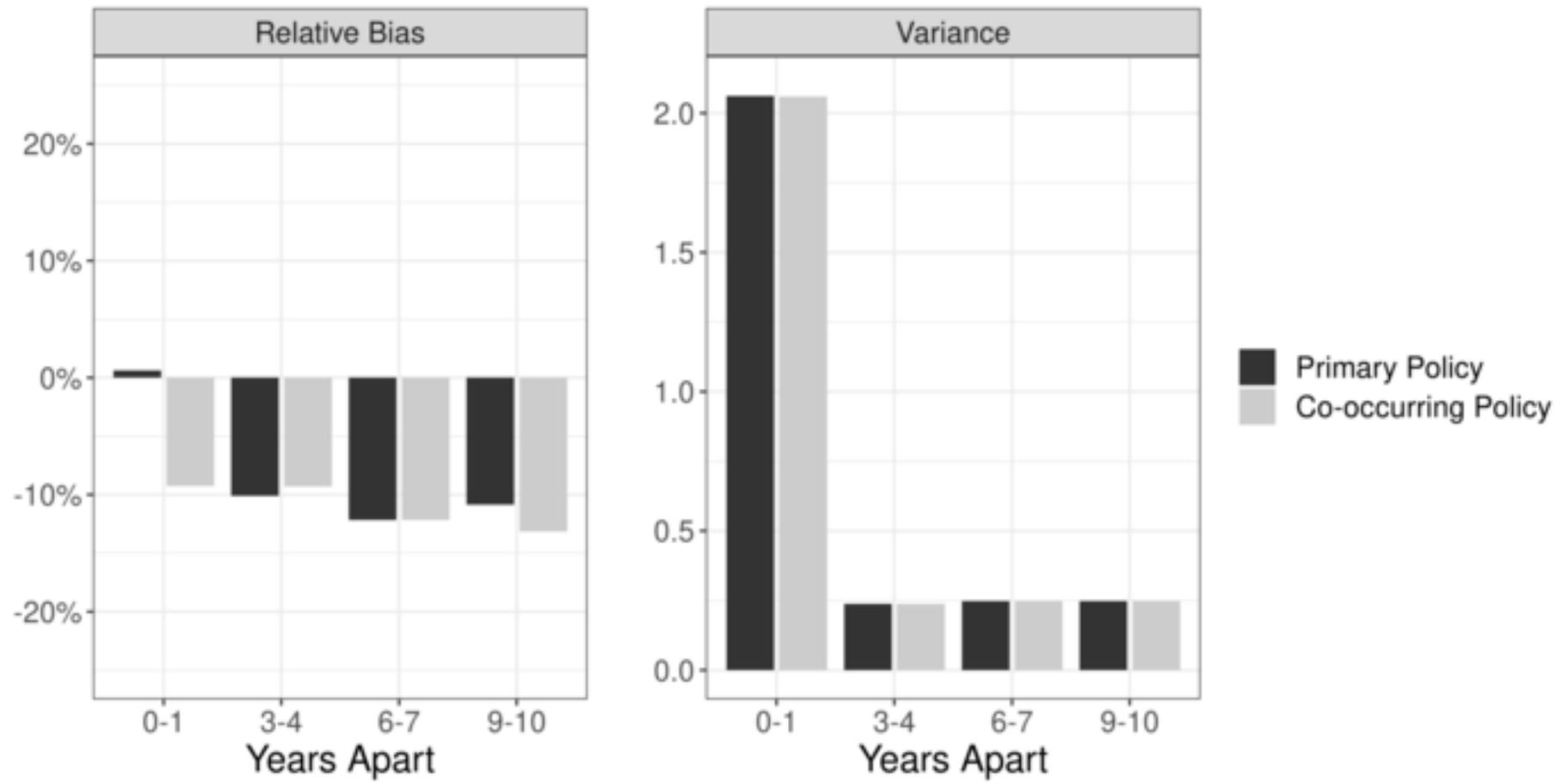

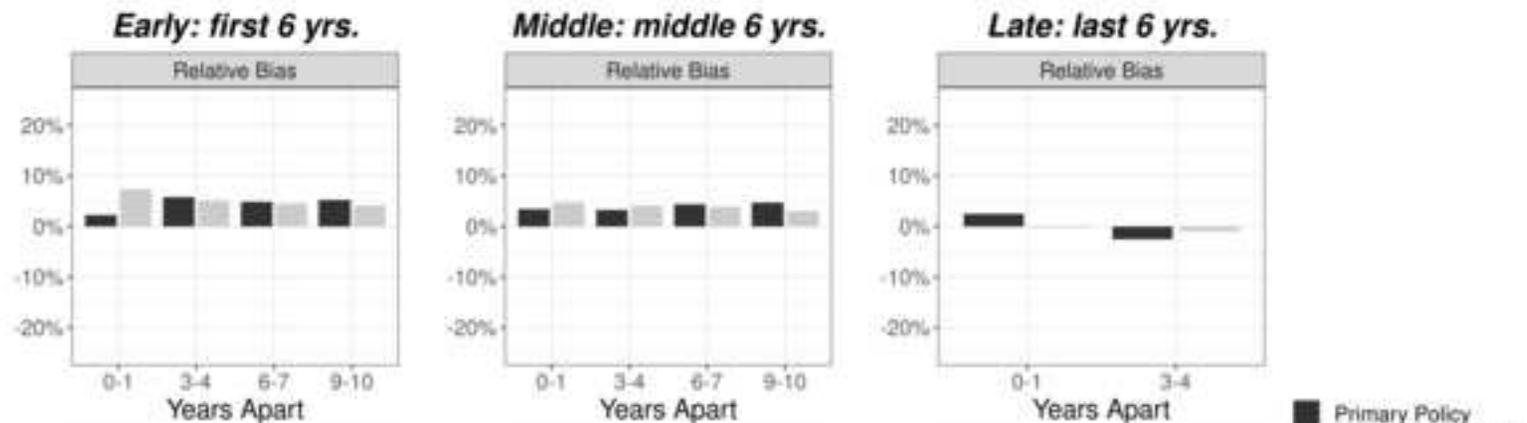

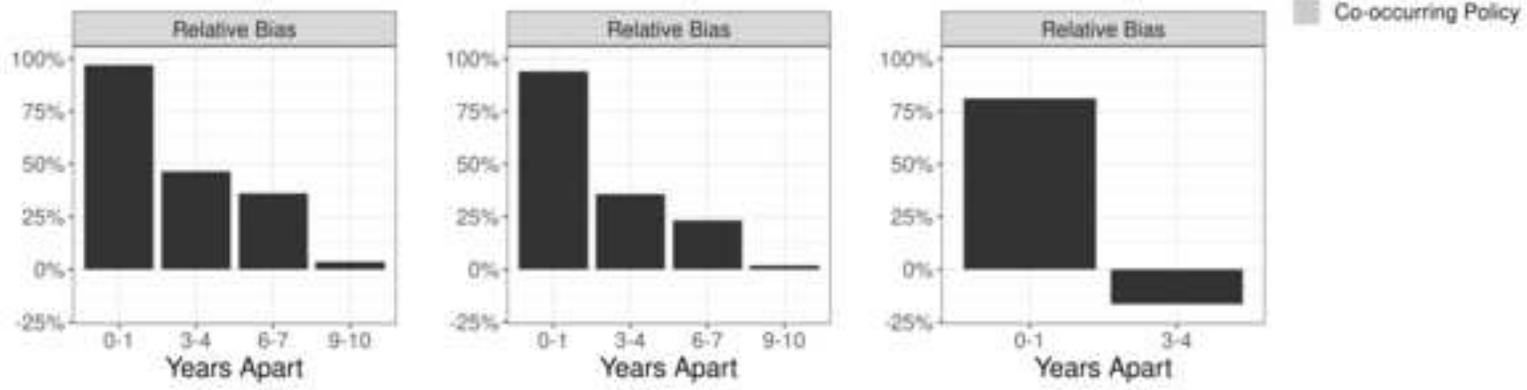

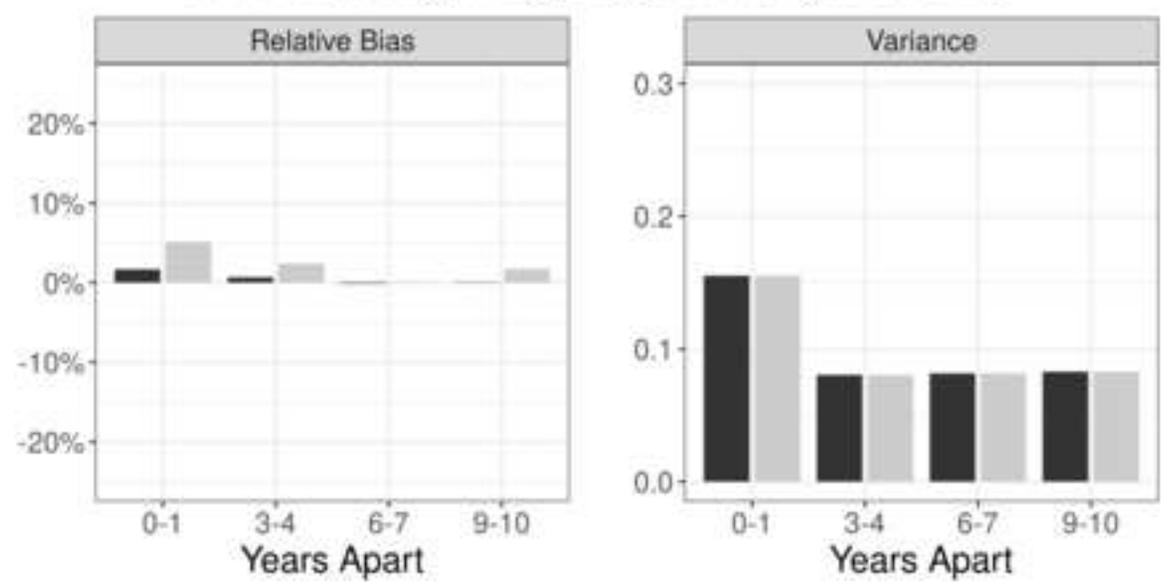

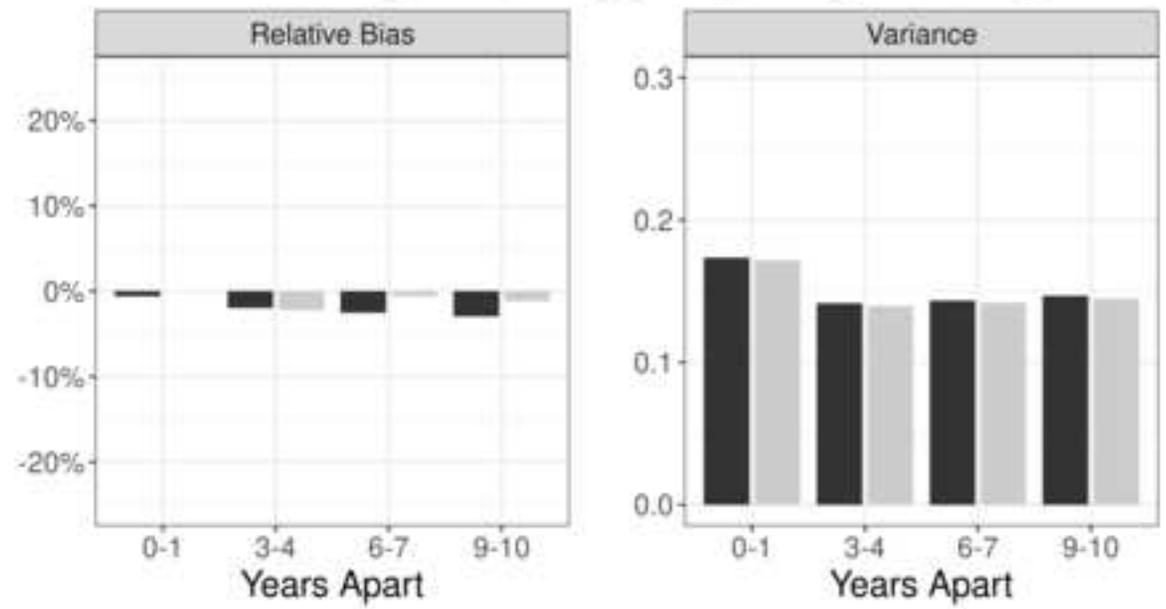